# Cross-concordances: terminology mapping and its effectiveness for information retrieval

**Philipp Mayr & Vivien Petras**

GESIS Social Science Information Centre (GESIS-IZ), Bonn, Germany
philipp.mayr|vivien.petras@gesis.org

**Abstract.** The German Federal Ministry for Education and Research funded a major terminology mapping initiative, which found its conclusion in 2007. The task of this terminology mapping initiative was to organize, create and manage 'cross-concordances' between controlled vocabularies (thesauri, classification systems, subject heading lists) centred around the social sciences but quickly extending to other subject areas. 64 crosswalks with more than 500,000 relations were established. In the final phase of the project, a major evaluation effort to test and measure the effectiveness of the vocabulary mappings in an information system environment was conducted. The paper reports on the cross-concordance work and evaluation results.

## 1. Introduction

In Germany, an ambitious project for one-stop academic search is the vascoda portal[1], a joint project between the Federal Ministry for Education and Research and the German Research Foundation. Vascoda provides a common search interface for a multitude of disciplinary and interdisciplinary databases (e.g. indexing and abstracting services, library catalogs, current content databases, full-text article databases, etc.) and internet resource collections (see Depping, 2007 for an overview). Since 2007, vascoda is partner in the global science gateway WorldWideScience.org[2].

Key concept of the vascoda portal is to structure and integrate high-quality information sources from more than 40 providers in one search space (approx. 81 million documents). The search space is organized in distributed disciplinary portals ("virtual subject libraries") and each of the integrated collections is clustered in the vascoda subject groups (Engineering and Physical Sciences; Medicine and Life Sciences; Law, Economics and Social Sciences; Humanities; Regions / Cultural Areas; Multidisciplinary Collections).

The vascoda portal contains many information collections that are meticulously developed and structured. They have sophisticated subject metadata schemes (subject headings, thesauri or classifications) to describe and organize the content of the documents on an individual collection level. The general search interface, however, only provides a free-text search over all metadata fields without regarding the precise subject access tools that were originally intended for these information collections. Because of the distributed nature of the collections and the multiple and varied subject access schemes, it is considered a difficult technical and managerial problem to integrate all these heterogeneous information sources with the same powerful but detailed subject access tools in one search interface.

At the same time, large-scale web search applications add both new collections and advanced subject access features to their expanding repertoire (e.g. clustering and video search in

---

[1] http://www.vascoda.de
[2] http://worldwidescience.org/

Flickr). Another prominent example are semantic web applications[3], which derive advanced reasoning functions from integrating ontologies and other semantic data. If large-scale contemporary information organization efforts like the semantic web strive to provide more structure and semantic resolution of information content, how is it possible that advanced interfaces for digital libraries scale back on exactly the same issue?

In 2004, the German Federal Ministry for Education and Research funded a major terminology mapping initiative (the KoMoHe project[4]) at the GESIS Social Science Information Centre in Bonn (GESIS-IZ), which found its conclusion at the end of 2007. One task of this initiative was to organize, create and manage 'cross-concordances' (cross-mappings) between major controlled vocabularies centered around the social sciences but quickly extending to other subject areas. The main objective of the project was to establish, implement and evaluate a terminology network to enable semantic integration of heterogeneous resources for search in a typical digital library environment.

The goal of semantic integration is to connect different information systems through their subject metadata - enabling distributed search over several information systems together with the advanced subject access tools provided by the individual databases. Through the mapping of different subject terminologies, a "semantic agreement" for the overall collection to be searched on is achieved. Terminology mapping – the mapping of words and phrases of one controlled vocabulary to the words and phrases of another – enables the seamless switch from a one-database-search to distributed search scenarios in the digital library world.

This paper describes the terminology mapping project KoMoHe and involved vocabularies and databases, the implementation of the developed cross-concordances in search as well as the results and findings of an extensive information retrieval evaluation analyzing the impact of terminology mappings on recall and precision in search.

## 2. Semantic heterogeneity

Generally there are two main approaches to treat semantic heterogeneity in digital libraries: Intellectual and automatic approaches. Essential for all efforts in terminology mapping is the acceptance of the remaining and unchangeable discrepancy between different terminologies. None of them can solely be responsible for the transfer burden between heterogeneous collections, mainly because of quality and costs reasons. Important is that the approaches bilaterally operate on the database level. According to Krause (2003) the approaches should be complemented by one another and work together.

- Cross-concordances between controlled vocabularies: The different concept systems are analyzed in a user context and an attempt made to relate intellectually their conceptualization. This idea should not be confused with the construction of metathesauri. While establishing cross-concordances, there is no attempt made to standardize existing concept worlds. Cross-concordance encompasses only partial union of existing terminological systems. They cover with it the static remaining part of the transfer problematic. Such concordances mostly offer mappings (see Table 1 and 2) in the sense of synonym or similarity/hierarchy relations but also as a deductive rule relation.

---

[3] See http://www.w3.org/2001/sw/ for a starting point.
[4] http://www.gesis.org/en/research/information_technology/komohe.htm

- Quantitative-statistical approaches: The transfer problem can be generally modeled as a fuzzy problem between two content description languages. For the vagueness addressed in information retrieval between terms e.g. within the user inquiry and the data collections, different automatic operations have been suggested (probability procedures, fuzzy approaches and neuronal networks) that can be used on the transfer problematic (Hellweg et al., 2001). The individual document can be indexed into individual documents in two concept schemata or whereby two different and differently indexed documents can be put in some relation to each other. Procedures of these types need training data. For the multilingual IR the same text can be in two languages.

When treatment of semantic heterogeneity (e.g. cross-concordances) is implemented in a distributed search scenario, a system of varied information collections can be searched with the subject metadata scheme one is familiar with. Terminology mappings could support distributed search in several ways. First and foremost, they should enable seamless search in databases with different subject metadata systems. Additionally, they can serve as tools for vocabulary expansion in general since they present a vocabulary network of equivalent, broader, narrower and related term relationships (see term examples in Table 1 and 2). Thirdly, this vocabulary network of semantic mappings can also be used for query expansion and reformulation.

Not only is the query being formulated in precise search statements but the terminology mapping service automatically translates the query in all the different terminologies implemented by the other information collections assembled in the digital library. A searcher can seamlessly switch between different information resources because the semantic translation between different terminologies used is done automatically.

For interdisciplinary information systems, semantic integration not only increases the success chances for distributed searches over collections with different subject metadata schemes but it also provides a window into a different disciplinary framework and domain-specific language for the searcher, if the mapped vocabularies are made available (see e.g. Figure 1).

Semantic mappings additionally play a big role in providing a transfer methodology between foreign-language databases. As a mapping can be created between controlled vocabularies from different databases or different disciplines, so can the mapping also provide a translation in the traditional sense: for example in table 1 from a German terminology to an English one.

Table 1 presents two seed terms (left column) in the German Thesaurus for the Social Sciences (TheSoz) and intellectually related end terms in mapped vocabularies (end vocabularies). Relationship types between terms are further explained in table 2.

| Start term TheSoz | Relation | End term | End vocabulary |
|---|---|---|---|
| Weiterbildung engl: "further education" | = | Weiterbildung | Psyndex, STW, Infodata, SWD, BISp, DZI |
| | ^ | Berufsfortbildung | FES |
| | = | Further education | CSA-ASSIA |
| | = | Continuing education | CSA-PEI |
| | = | Adult Education | CSA-SA |
| | < | Education | CSA-WPSA |
| | = | Erwachsenenbildung | IBLK |

| Start term TheSoz | Relation | End term | End vocabulary |
|---|---|---|---|
| Meinungsforschung engl: "opinion research" | 0 | | Psyndex |
| | ^ | Einstellungsforschung | IAB |
| | = | Opinion Polls | CSA-ASSIA |
| | = | Opinions + Research | CSA-SA |
| | < | Research | CSA-PEI |
| | = | Public Opinion Research | CSA-WPSA |
| | = | Public Opinion Polls | ELSST |
| | = | Meinungsumfrage/Meinungs-forschung | IBLK |

Table 1. Start or seed terms in the TheSoz vocabulary and a selection of end terms (semantic mappings).

In recent years, different organizations have initiated efforts to provide semantic integration for information systems. In the United States, OCLC launched the Terminology Services[5] project (Vizine-Goetz, 2004, 2006) to offer web services for terminology mappings between different (mostly Library of Congress) controlled vocabularies like the DDC, LCC, LCSH or MeSH. In Europe, the Delos2 Network of Excellence in Digital Libraries programme devoted one work package (WP5) to the problem of knowledge extraction and semantic operability (Patel et al., 2005). Another report commissioned by JISC provides an overview of terminology services with a focus on the UK efforts (Tudhope, Koch et al., 2006). Other projects are the CRISSCROSS[6] project at German National Library and Cologne University of Applied Sciences which creates a multilingual, thesaurus-based research vocabulary between the Subject Heading Authority Files (SWD) and notations of the Dewey Decimal Classification (DDC) (see Panzer, 2008). The Agricultural Information Management Standards department at the FAO[7] is involved in various terminology mapping initiatives (e.g. Liang & Sini, 2006). The High-Level Thesaurus Project (HILT[8]) at the University of Strathclyde is another example for a project with long-term experience in developing terminology mapping technologies (Macgregor et al., 2007).

## 3. Terminology mapping approach at GESIS-IZ

Semantic interoperability can be achieved in different ways. For a good overview of different terminology mapping methodologies and mapping projects, see Zeng & Chan (2004, 2006a, 2006b), Doerr (2001, 2004), and Hellweg et al. (2001).

---

[5] http://www.oclc.org/research/projects/termservices/
[6] http://www.d-nb.de/eng/wir/projekte/crisscross.htm
[7] http://www.fao.org/aims/
[8] http://hilt.cdlr.strath.ac.uk/

The project KoMoHe focused on cross-concordances. We define cross-concordances as intellectually (manually) created crosswalks that determine equivalence, hierarchy, and association relations between terms from two controlled vocabularies.

Typically, vocabularies will be related bilaterally, that is, a cross-concordance relating terms from vocabulary A to vocabulary B as well as a cross-concordance relating terms from vocabulary B to vocabulary A are established. Bilateral relations are not necessarily symmetrical. For example, the term 'Computer' in system A is mapped to the term 'Information System' in system B, but the same term 'Information System' in system B is mapped to another term 'Data base' in system A.

Our approach allows the following 1:1 or 1:n relations:
- Equivalence (=) means identity, synonym, quasi-synonym
- Hierarchy (Broader terms <; narrower terms >)
- Association (^) for related terms
- An exception is the Null (0) relation, which means that a term can't be mapped to another term (see mapping number 4 in Table 2).

In addition, every relation must be tagged with a relevance rating (high, medium, and low). The relevance rating is a secondary but weak instrument to adjust the quality of the relations. They are not used in our current implementations.

Table 2 presents typical unidirectional cross-concordances between two vocabularies A and B.

| No | Vocabulary A | Relation | Vocabulary B | Description |
|---|---|---|---|---|
| 1 | hacker | = | hacking | Equivalence relationship |
| 2 | hacker | ^+ | computers + crime | 2 association relations (^) to term combinations (+) |
| 3 | hacker | ^+ | internet + security | |
| 4 | isdn device | 0 | | Null-relation. Concept can't be mapped, term is too specific. |
| 5 | isdn | < | telecommunications | Narrower term relationship |
| 6 | documentation system | > | abstracting services | Broader term relationship |

Table 2. Cross-concordance examples (unidirectional).

The mappings in the KoMoHe project involve all or major parts of the vocabularies. Vocabularies were analyzed in terms of topical and syntactical overlap before the mapping started. All mappings are created by researchers or terminology experts. Essential for a successful mapping is an understanding of the meaning and semantics of the terms and the internal relations of the concerned vocabularies. This includes syntactic checks of word stems but also semantic knowledge to look up synonyms and other related terms.

The mapping process is based on a set of practical rules and guidelines (see e.g. Patel et al., 2005). During the mapping of the terms, all intra-thesaurus relations (including scope notes) are consulted. Recall and precision of the established relations have to be checked in the associated databases. This is especially important for combinations of terms (1:n relations). One-to-one (1:1) term relations are preferred. Word groups and relevance adjustments have to be made consistently.

In the end, the semantics of the mappings are reviewed by experts and samples are empirically tested for document recall and precision. All things considered, it is a qualitative but cost-intensive and time-consuming effort to generate a terminology network solely with cross-concordances.

## 3.1 Results of the mapping initiative

To date, 25 controlled vocabularies from 11 disciplines and 3 languages (German, English and Russian) have been connected with vocabulary sizes ranging from 1,000 – 17,000 mapped terms per vocabulary (see Figure 2 for a detailed view of the mappings). More than 513,000 relations were generated in 64 crosswalks (30 bilateral[9] and 4 unidirectional cross-concordances). Figure 1 depicts the established network of cross-concordances by discipline.

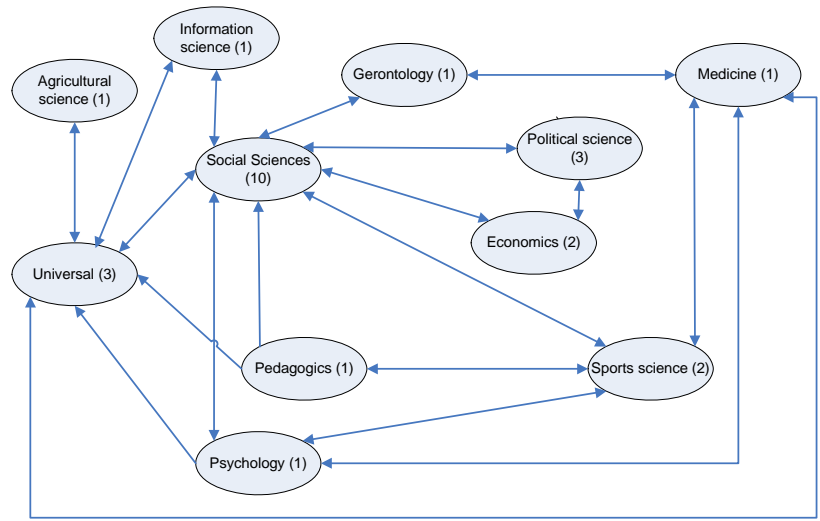

Figure 1. Network of terminology mappings in the KoMoHe project. The numbers in brackets contain the number of mapped controlled vocabularies in a discipline.

The project generated cross-concordances between the following controlled vocabularies (thesauri, descriptor lists, classifications, and subject headings) which all play a role in the subject specific collections of vascoda. Several cross-concordances from the previous projects CARMEN[10] and infoconnex[11] were incorporated.

The vocabularies involved in the project KoMoHe are mostly in German, English (N=8), Russian (N=1), or multilingual (e.g. AGROVOC, IBLK, DDC). Some vocabularies have English or German translations of terms (e.g. THESOZ, PSYNDEX, MESH, INION, STW).

Mapped thesauri (N=16):
- AGROVOC Thesaurus (AGROVOC): A vocabulary in the *agricultural* domain which contains round 39,000 terms. Mapping to: SWD.
- CSA Thesaurus Applied Social Sciences Index and Abstracts (CSA-ASSIA): A vocabulary in the *social science* domain which contains round 17,000 terms. Mapping to: THESOZ.
- CSA Thesaurus PAIS International Subject Headings (CSA-PAIS): A vocabulary in the *political science* domain which contains round 7,000 terms. Mapping to: IBLK.
- CSA Thesaurus Physical Education Index (CSA-PEI): A vocabulary in the *sports science* domain which contains round 1,800 terms. Mapping to: THESOZ.
- CSA Thesaurus of Political Science Indexing Terms (CSA-WPSA): A vocabulary in the *social and political science* domain which contains round 3,100 terms. Mapping to: THESOZ.
- European Language Social Science Thesaurus (ELSST): A vocabulary in the *social science* domain which contains round 3,200 terms. Mapping to: THESOZ.
- INFODATA Thesaurus (INFODATA): A vocabulary in the *information science* domain which contains round 1,000 terms. Mapping to: THESOZ and SWD.

---

[9] A bilateral cross-concordance is counted as two crosswalks.
[10] http://www.bibliothek.uni-regensburg.de/projects/carmen12/index.html.en
[11] http://www.infoconnex.de/

- Psyndex Terms (PSYNDEX): A vocabulary in the *psychological* domain which contains round 5,400 terms. Mapping to: THESOZ, SWD, BISP, MESH and BILDUNG.
- Standard Thesaurus Wirtschaft (STW): A vocabulary in the *economics* domain which contains round 5,700 terms. Mapping to: THESOZ, SWD, IAB and IBLK.
- Thesaurus Bildung (BILDUNG): A vocabulary in the *pedagogic* domain which contains round 50,000 terms. Mapping to: THESOZ, SWD, PSYNDEX and BISP.
- Thesaurus Internationale Beziehungen und Länderkunde (IBLK): A vocabulary in the *political science* domain which contains round 8,400 terms. Mapping to: THESOZ, STW, TWSE and CSA-PAIS.
- Thesaurus Sozialwissenschaften (THESOZ): A vocabulary in the *social science* domain which contains round 7,700 terms. Mapping to: GEROLIT, DZI, FES, CSA-WPSA, CSA-ASSIA, CSA-SA, CSA-PEI, ELSST, IAB, IBLK, STW, SWD, BILDUNG, PSYNDEX, INFODATA and BISP.
- Thesaurus für wirtschaftliche und soziale Entwicklung (TWSE): A vocabulary in the *political science* domain which contains round 2,800 terms. Mapping to: IBLK.
- Thesaurus of Sociological Indexing Terms (CSA-SA): A vocabulary in the *social science* domain which contains round 4,300 terms. Mapping to THESOZ.
- Thesaurus of the Deutschen Instituts für soziale Fragen (DZI): A vocabulary in the *social science* domain which contains round 1,900 terms. Mapping to THESOZ.
- Thesaurus of the Deutschen Zentrums für Altersfragen (GEROLIT): A vocabulary in the *gerontology* domain which contains round 1,900 terms. Mapping to THESOZ and MESH.

Mapped descriptor lists (N=4):
- Descriptors of the Bundesinstitut für Sportwissenschaft (BISP): A vocabulary in the *sports science* domain which contains round 7,400 terms. Mapping to THESOZ, MESH and BILDUNG.
- Descriptors of the Friedrich-Ebert Stiftung (FES): A vocabulary in the *social science* domain which contains round 4,000 terms. Mapping to THESOZ.
- Descriptors of the Institut für Arbeitsmarkt- und Berufsforschung (IAB): A vocabulary in the *social science* domain which contains round 6,800 terms. Mapping to THESOZ and STW.
- Descriptors of the Institute of Scientific Information on Social Sciences of the Russian Academy of Sciences (INION): A vocabulary in the *social science* domain which contains round 7,000 terms. Mapping to THESOZ.

Mapped classifications (N=3):
- Dewey Decimal Classification (DDC): An *univeral* vocabulary which contains thousands of notations. Mapping to RVK.
- Journal of Economic Literature Classification System (JEL): A vocabulary in the *economics* domain which contains round 1,000 notations. Mapping to STW.
- Regensburger Verbundklassifikation (RVK): An *univeral* vocabulary which contains thousands of notations. Mapping to DDC.

Mapped subject heading lists (N=2):
- Medical Subject Headings (MESH): A vocabulary in the *medicine* domain which contains round 23,000 terms. Mapping to PSYNDEX, GEROLIT, BISP and SWD.
- Schlagwortnormdatei (SWD): An *universal* vocabulary which contains round 650,000 terms. Mapping to THESOZ, MESH, STW, AGROVOC and INFODATA.

Figure 2 gives an overview of all 64 crosswalks. The Thesaurus Sozialwissenschaften (THESOZ) is the vocabulary with the most incoming and outgoing mappings and due to its centrality the THESOZ is displayed in the middle of the net. Other vocabularies like SWD or PSYNDEX play central roles for switching into other domains. The mapping DDC-RVK is the only cross-concordance which is not connected. Possibly, the terminology work done by the project CRISSCROSS which maps SWD to DDC could be utilized to connect this disconnected pair. The mapping JEL-STW is one example for a unidirectional (one-way) cross-concordance from JEL to STW.

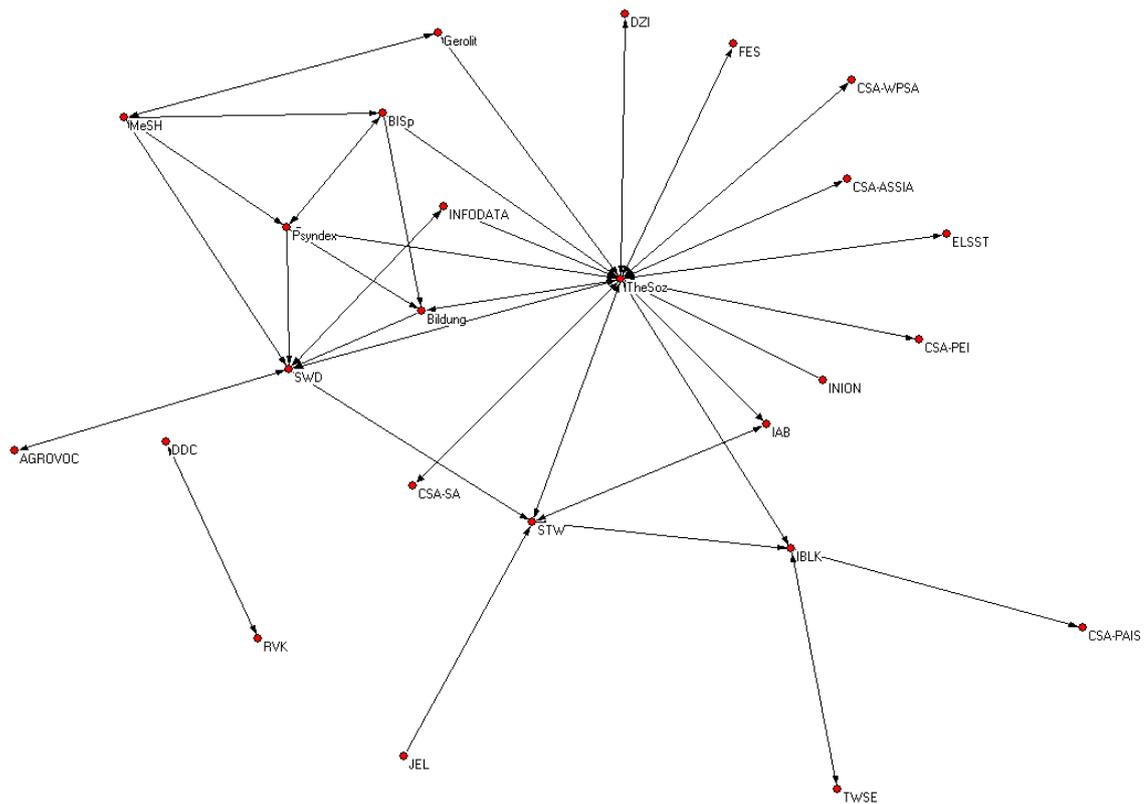

Figure 2. Net of mapped vocabularies in the KoMoHe project.

The 513,000 relations available in our cross-concordance database involve more than 181,000 unique searchable concepts (unique controlled descriptors or descriptor combinations or notations). In average (per cross-concordance) 6,500 start terms are mapped to 3,600 terms in the end vocabulary (1.2 relations per term).

Figure 3 displays the distribution of relationship types in the project (compare Table 2). Equivalence relation (round 45%) is the most frequent relationship type between terms. Just 12% of all relations are 'Null relations' (no mapping of a term possible).

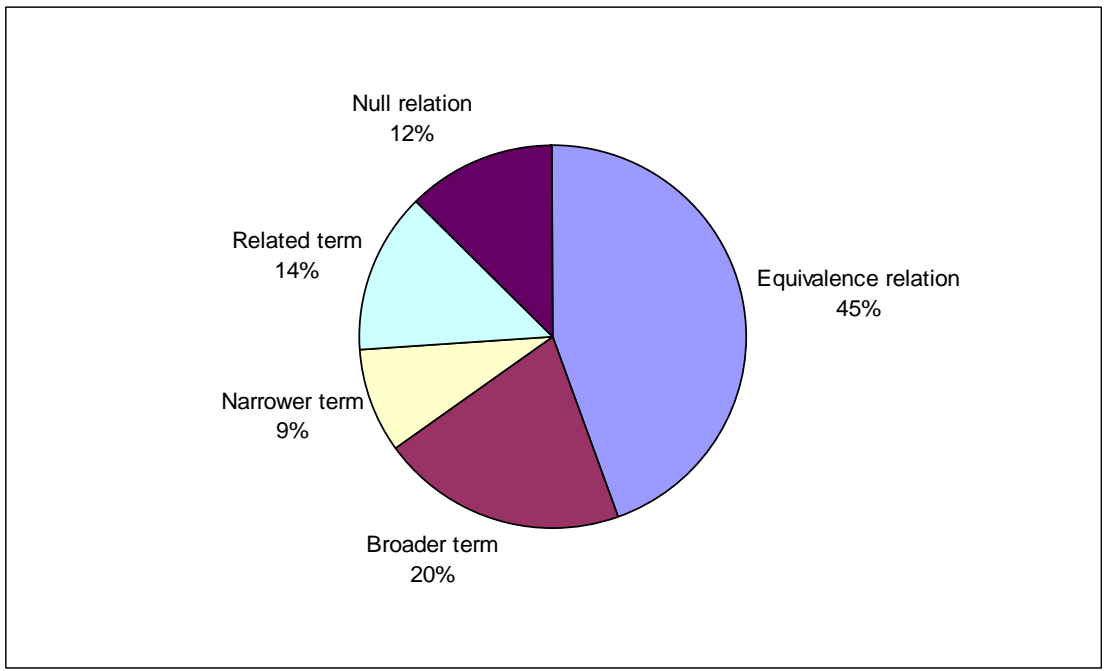

Figure 3. Distribution of relationship types across all cross-concordances.

## 3.2 Implementation of the Cross-concordances

A relational database was created to store the cross-concordances for later use. It was found that the relational structure is able to capture the number of different controlled vocabularies, terms, term combinations, and relationships appropriately. The vocabularies and terms are represented in list form, independent from each other and without attention to the syndetic structure of the involved vocabularies. Orthography and capitalization of controlled vocabulary terms were normalized. Term combinations (i.e. computers + crime as related combination for the term hacker) were also stored as separate concepts.

To search and retrieve terminology data from the database, a web service (called heterogeneity service or HTS in Figure 4, see Mayr & Walter, 2008) was built to support cross-concordance searches for individual start terms, mapped terms, start and destination vocabularies as well as different types of relations. One implementation, which uses the equivalence relations, looks up search terms in the controlled vocabulary term list and then automatically adds all equivalent terms from all available vocabularies to the query. If the controlled vocabularies are in different languages, the heterogeneity service also provides a translation from the original term to the preferred controlled term in the other language. If the original query contains a Boolean command, it remains intact after the query expansion (i.e. each query word gets expanded separately). Because of performance issues, the cross-concordance query expansion doesn't distinguish between different databases and their preferred controlled vocabulary terms given a concept, but adds all equivalent terms to the query. In principle, this use of the terminology network expands a query with synonyms or quasi-synonyms of the original query terms.

# 4. Cross-concordance evaluation

## 4.1 General Questions

Although the need for terminology mappings is generally acknowledged by the community and many mapping projects are undertaken, the actual effectiveness and usefulness of the project outcomes is rarely evaluated stringently. Many questions can be asked of the terminology networks created in these mappings, e.g.:
- How many expressions can be found for a concept?
- Which concepts are related?
- Are vocabularies broader or narrower in scope?
- Which terminologies are very similar to each other?
- Which disciplines / subject are adjacent or far apart?
- How much overlap exists between different databases or controlled vocabularies in a particular subject?

The most important question, and the one most mappings are created for, is how effective and helpful the mapping are in an actual search. In an information portal with many different databases, the question becomes crucial whether cross-concordances can enable a distributed search. Can they bridge the differences in language in order to facilitate a seamless search with the same query across different databases?

When evaluating terminology mappings, the analytical starting point is critical. What is examined: the quality of the mappings per se or the quality of the associated search? The quality of the mappings is a prerequisite for an improved quality of search. For the cross-concordances in the KoMoHe project, every mapping was checked by subject experts from the partnering institutions. The manual creation and careful checks provides assurance that the mappings are sensible, appropriate and of consistent quality.

The intrinsic features of cross-concordances (and their impact on search) can differ depending on the mapped controlled vocabularies and on external factors in the cross-concordance creation process. For example, the creation date of the cross-concordance can affect the number of relations per starting term. Earlier in the project, fewer relations were formed. Cross-concordances from an earlier project (CARMEN) were discussed among a group of experts and are more selective. Changes in the controlled vocabularies or indexing practices can also impact the quality of the cross-concordance. Other differences can be observed in:
- Sizes of start / destination terminologies
- Differences in pre- and post-coordination vocabularies
- Number of relations
- Number of mapped destination terms (coverage / overlap)
- Distribution of relations (equivalence, broader term, narrower term, related term, null relation)
- Distribution of relevances (high, medium, low)
- Identical term mappings
- Disparity in specificity (e.g. vocabularies that are very broad or very narrow in scope)
- Combination of mapped terms (mapping consists of more than one end term)

A quantitative analysis can give some insight into the basic features of a cross-concordance, but it can not determine the quality improvements gained from using specific mappings in search. We have devised an information retrieval test with the goal of evaluating the application of cross-concordances in a real-world search scenario.

## 4.2 Information Retrieval Test Design

In search, several factors come into play when evaluating the quality of the terminology mappings: the cross-concordances themselves, but also the contents of the involved databases, their coverage or overlap of contents, the search interface, or the retrieval ranking utilized. The goal was to evaluate the impact of the cross-concordances, the actual retrieval conditions (interface, ranking method, etc.) were therefore kept as stable as possible.

The basic idea for utilizing cross-concordances is to translate the search terms into other terminologies to facilitate the search across different databases and terminologies. Leveraging the cross-concordances should expand the search space, correct ambiguities and imprecision in the query formulation and therefore find more relevant documents for a given query.

The application of cross-concordances in search also presents a caveat: they could impact the speed and ease-of-use of the search process itself. One premise for the technical implementation of terminology mappings should be that they are utilized inconspicuously for the searcher. The mappings should improve the search experience without increasing the effort for the user of the information system. By using a strictly automatic approach for the input of cross-concordances during the evaluation, no manual interference in the form of human query re-formulation was necessary.

Two information retrieval tests were devised to evaluate the quality of the cross-concordances in search:

**Test 1:** *Does the application of term mappings improve search over a non-transformed subject (i.e. controlled vocabulary) search?*

In test 1, a query was translated into the terms of a controlled vocabulary (A) and then searched against the controlled term fields of a bibliographic database with a different controlled vocabulary (B). The search was repeated with the help of the cross-concordance A→B, translating the original controlled vocabulary search terms into the controlled vocabulary terms for the destination database. Figure 4 shows a graphical representation of the process. The retrieval results were compared.

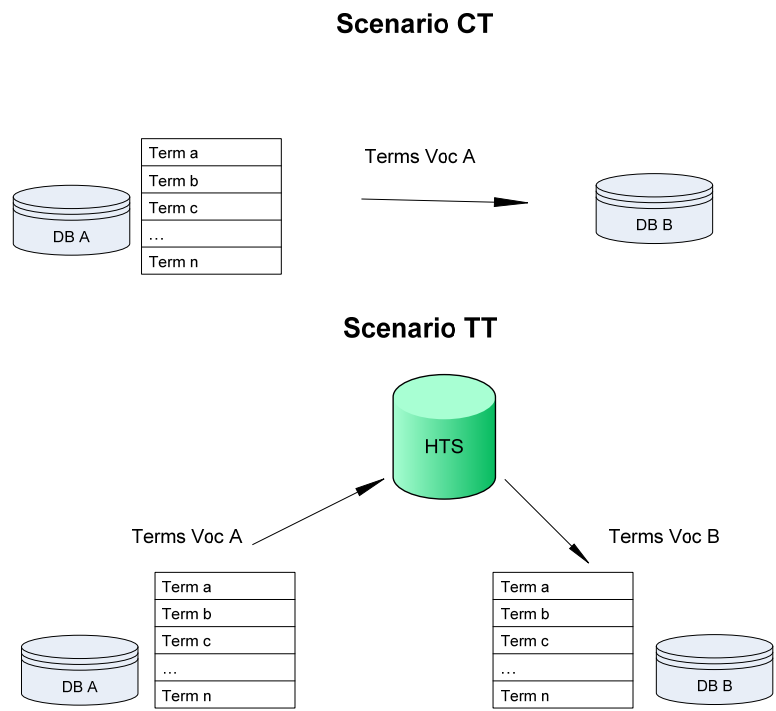

Figure 4. Cross-concordance information retrieval evaluation set-up. Scenario CT = controlled term search. Scenario TT = term-transformed search. HTS = heterogeneity service.

If the results of the two searches are the same, the application of cross-concordances for term transformation does not have an effect. If the search results deteriorate, cross-concordances have a negative effect, and if the search results improve, which is expected, the use of cross-concordances has a positive effect on the search.

**Test 2:** *Does the application of term mappings improve for a free-text search?*

In test 2, the original query was used in a free-text search scenario (most searchers do not use the appropriate controlled vocabulary). A free-text search searches the query terms not only in the controlled vocabulary fields, but also in the title and abstract fields. In the experiment, the original query was first searched in the free-text fields of the database. Secondly, the query terms were looked up in the cross-concordance if a term mapping existed. The new terms from the cross-concordance were added to the original query.

An example illustrates the differences between tests 1 and 2: a natural language query asks for documents about 'family relations'. 'Family relations' is already a controlled vocabulary term in vocabulary A and doesn't need to be translated into controlled terms, so it is used for the first search against the controlled term fields in database B: *Test 1 CT: Family relations.*
The cross-concordance A→B maps the phrase 'family relations' from vocabulary A to the term combination 'family' AND 'social relations' in vocabulary B. The second search in database B is therefore: *Test 1 TT: Family AND social relations.*

For test 2, the original query is searched in the free-text fields of database B (title, abstracts and controlled terms): *Test 2 FT: Family relations.*

Since the query terms occur in vocabulary A, a term mapping for the query can be found in the cross-concordance. The terms are added to the original query and searched against the free-text fields in database B: *Test 2 FT+TT: Family relations OR (Family AND social relations)*.

Test 1 only searches the controlled term fields, whereas test 2 also searches other fields where the query terms could occur by chance (title & abstract). Test 2 is generally a weaker test, since the mapped terms are appended to the query and do not replace the original query as in test 1. Since test 2 searches also the fields test 1 searches, test 2 could be considered to subsume test 1. In test 2, however, fewer term additions take place, because not all query terms occur in the original controlled vocabulary and fewer term mappings are found.

For query creation, the help of the producers or hosts of the tested databases were solicited to assure that realistic queries were developed. They were asked to provide 3-10 queries (average: 6-7) from their daily experience, which were translated into the controlled vocabularies of the tested databases. The natural-language free-text queries contained around 1-3 terms per queries, whereas the Boolean queries for controlled term search contained around 2-6 terms. For all term mappings in the information retrieval experiments, only the equivalence relations were used. All documents the information system listed were retrieved until a cut-off number of 1,000 ranked documents. Finally, the result sets of documents in each experiment were assessed for relevance to the question.

To evaluate the effect of cross-concordances, the classical information retrieval measures recall and precision based on the relevance assessments of the retrieved documents were used. The following measures were analyzed:
- Retrieved: average number of retrieved documents (across all search types)
- Relevant: average number of relevant retrieved documents (across all search types)
- Rel_ret: average number of relevant retrieved documents for a particular search type
- Recall: proportion of relevant retrieved documents out of all relevant documents (averaged across all queries of one search type)
- Precision: proportion of relevant retrieved documents out of all retrieved documents (averaged across all queries of one search type)
- P10: Precision at 10 = Precision after 10 retrieved documents
- P20: Precision at 20 = Precision after 20 retrieved documents

P10 and P20 were calculated in order to represent a realistic search scenario where a user commonly does not look past the first or second results page. For retrieval systems, which do not rank but list results by year or author, P10 and P20 are not meaningful.

**4.3 Cross-concordances and databases tested**

For both experiments, the cross-concordances were divided by discipline (intra- or interdisciplinary) and by language (mono- or bilingual). Intradisciplinary cross-concordances span vocabularies mostly in the social sciences area as most cross-concordances created in the project are situated in that discipline. The interdisciplinary cross-concordances are mapping vocabularies in the fields of economics, medicine, political science, psychology, and the social sciences. The monolingual cross-concordances included vocabularies in the German language; the bilingual cross-concordances included a German and an English vocabulary. Table 3 gives an overview over the number of tested cross-concordances per experiment:

| Test 1: Controlled term search | |
| --- | --- |
| Intradisciplinary cross-concordances | 5 (1 bilingual) |
| Interdisciplinary cross-concordances | 8 |
| Test 2: Free-text search | |
| Intradisciplinary cross-concordances | 6 (1 bilingual) |
| Interdisciplinary cross-concordances | 2 |

Table 3. Number of cross-concordances tested

The background behind separating the cross-concordances in this fashion is the hypothesis that cross-concordances between vocabularies in the same discipline (intradisciplinary) will overlap and contain more identical terms and therefore have a lesser effect on the retrieval results than the interdisciplinary cross-concordances. Term mappings between vocabularies in a different natural language (i.e. English → German) or between different notation systems (i.e. DDC → LCC) will also have a much bigger impact because the occurrences of identical terms or an overlap are more improbable.

For the experiments, bibliographic databases that contain between 70,000 – 16 million documents were included, most of them produced and hosted in Germany. Among them were abstracting and indexing databases, but also library catalogs. Table 4 gives an overview over the tested databases and their associated vocabularies:

| Vocabulary | Discipline | Database | Documents in DB |
| --- | --- | --- | --- |
| TheSoz – Thesaurus Sozialwissenschaften (GESIS-IZ) | Social Sciences | SOLIS | 345,086 |
| DZI – Thesaurus des Deutschen Instituts für soziale Fragen | Social Sciences | SoLit | 151,925 |
| SWD – Schlagwortnormdatei | General (Social Sciences Excerpt) | USB Köln Sowi OPAC | 72,729 |
| CSA – Thesaurus of Sociological Indexing Terms (Cambridge Scientific Abstracts) | Social Sciences | CSA Sociological Abstracts | 294,875 |
| Psyndex - Psyndex Terms | Psychology | Psyndex (ZPID) | Ca. 200,000 |
| STW – Standard Thesaurus Wirtschaft | Economics | Econis (ZBW Kiel) | Ca. 3,000,000 |
| IBLK - Thesaurus Internationale Beziehungen und Länderkunde (Euro-Thesaurus) | Political Science | World Affairs Online WAO (SWP Berlin) | 643,420 |
| Mesh – Medical Subject Headings | Medicine | Medline (Dimdi) | Ca. 16,800,000 |

Table 4. Vocabularies and databases in the KoMoHe IR test

Many cross-concordances and their respective databases could be tested in-house by indexing the documents with the open-source information retrieval system Solr[12] using the same processing and ranking modules for every database. For the databases not available in-house, the hosts were asked to provide ranked results lists for pre-determined queries.

For most databases, the term mappings were tested in both directions, going from vocabulary A to B (A → B) as well as from vocabulary B to A (B → A). As they constitute different searches (different queries depending on the start vocabulary) and different databases, they are independent.

---

[12] http://lucene.apache.org/solr/

## 5. Results of the evaluation

### 5.1 Test 1: Controlled term search

Test 1 evaluated whether the replacement of a query with vocabulary A terms (CT) with controlled vocabulary terms from vocabulary B (transformation through term mapping) (TT) would improve retrieval in database B. If the term mapping is imprecise or ambiguous or the vocabularies overlap, then the translation from the original query to the mapped query could introduce noise into the query formulation, which could then impede on the quality of the search.

Table 5 gives an overview of the average results over all 13 tested cross-concordances. The last line shows the difference in percentage points between the search types:

|    | Retrieved | Relevant | Rel_ret | Recall | Precision | P10 | P20 |
|----|-----------|----------|---------|--------|-----------|-----|-----|
| CT | 156.5 | 144.8 | 42.0 | 0.3152 | 0.2214 | 0.1987 | 0.1748 |
| TT | 325.4 | 144.8 | 88.2 | 0.6047 | 0.3391 | 0.3052 | 0.2848 |
|    |       |       |      | **91.8%** | **53.2%** | **53.6%** | **62.9%** |

Table 5. Test 1 evaluation results for all cross-concordances (N=13)

The search utilizing term transformations doubles the number of retrieved documents, more documents containing the query terms are found. Recall increases by almost 100%, whereas precision increases by more than 50%. The use of a cross-concordance in this particular search finds not only more relevant documents (recall) but is still more accurate (precision) than a search without the term transformation.

However, this huge improvement is partly due to the translation between English and German in the bilingual cross-concordance. Whereas monolingual term mappings might be ineffective because the mapped terms are identical, this will not be the case in translated mapping. Table 6 show the retrieval results when the bilingual cross-concordance is removed from the test set:

|    | Retrieved | Relevant | Rel_ret | Recall | Precision | P10 | P20 |
|----|-----------|----------|---------|--------|-----------|-----|-----|
| CT | 169.6 | 141.2 | 45.5 | 0.3415 | 0.2399 | 0.2153 | 0.1894 |
| TT | 320.5 | 141.2 | 87.6 | 0.6113 | 0.3431 | 0.3126 | 0.2877 |
|    |       |       |      | **79.0%** | **43.1%** | **45.2%** | **51.9%** |

Table 6. Test 1 evaluation results for all monolingual cross-concordances (N=12)

Because of term overlap, the retrieval results should be different for cross-concordances spanning two disciplines (interdisciplinary) or cross-concordances within the same disciplinary area (intradisciplinary). If the test results are separated by disciplinarity, we can see significant changes in the retrieval results. For intradisciplinary cross-concordances, recall and precision increase but not as much. A smaller or negative change in precision should actually be expected as commonly in information retrieval precision and recall are in an inverse relationship with each other (if recall rises, precision falls).

Table 7 shows the average recall and precision measures for all and only the monolingual intradisciplinary cross-concordances. For monolingual intradisciplinary cross-concordances, precision and recall still increase but much less than for all cross-concordances.

|    | Retrieved | Relevant | Rel_ret | Recall | Precision | P10 | P20 |
|----|-----------|----------|---------|--------|-----------|--------|--------|
| CT | 126.6 | 101.3 | 36.2 | 0.3726 | 0.2491 | 0.2002 | 0.1637 |
| TT | 238.9 | 101.3 | 59.9 | 0.5189 | 0.3335 | 0.2784 | 0.2352 |
|    |       |       |      | **39.3%** | **33.9%** | **39.1%** | **43.7%** |
|    | Monolingual ||||||||
| CT | 158.2 | 79.7 | 45.3 | 0.4657 | 0.3113 | 0.2503 | 0.2046 |
| TT | 202.9 | 79.7 | 51.0 | 0.5174 | 0.3441 | 0.2939 | 0.2315 |
|    |       |       |      | **11.1%** | **10.5%** | **17.4%** | **13.2%** |

Table 7. Test 1 evaluation results for intradisciplinary cross-concordances (N=5)

An extraordinary improvement in recall and precision can be observed for the application of interdisciplinary cross-concordances. Recall and precision increase significantly more than for the average cross-concordance (see table 8):

|    | Retrieved | Relevant | Rel_ret | Recall | Precision | P10 | P20 |
|----|-----------|----------|---------|--------|-----------|--------|--------|
| CT | 175.2 | 171.9 | 45.6 | 0.2794 | 0.2041 | 0.1978 | 0.1817 |
| TT | 379.4 | 171.9 | 105.9 | 0.6583 | 0.3426 | 0.3220 | 0.3157 |
|    |       |       |      | **135.6%** | **67.8%** | **62.8%** | **73.7%** |

Table 8. Test 1 evaluation results for interdisciplinary cross-concordances (N=8)

Utilizing cross-concordances has more than a positive effect on the controlled term search. The result set is not only bigger but also more precise. The biggest impact can be observed for cross-concordances spanning more than one discipline.

**5.2 Test 2: Free-text search**

Test 2 evaluated whether adding controlled vocabulary terms gained from mapping natural language query terms to the controlled vocabulary of a database (FT-CK) to a free-text query (FT) would improve retrieval results. For some of the individual queries in the tests, no changes to the queries were made because no matching controlled vocabulary terms could be found. Table 9 shows the retrieval results for all 8 tested cross-concordances:

|       | Retrieved | Relevant | Rel_ret | Recall | Precision | P10 | P20 |
|-------|-----------|----------|---------|--------|-----------|--------|--------|
| FT    | 155.3 | 106.4 | 56.2 | 0.6026 | 0.4551 | 0.4101 | 0.3682 |
| FT-CK | 266.8 | 106.4 | 72.8 | 0.7273 | 0.3934 | 0.3203 | 0.3083 |
|       |       |       |      | **20.7** | **-13.6** | **-21.9%** | **-16.3%** |

Table 9. Test 2 evaluation results for all cross-concordances (N=8)

The results show that not only more but more relevant documents are found. Average recall still increases by 20%. Generally, controlled terms simply added to a query can still improve retrieval results. However, a drop in precision is observed, which is nevertheless not as big as the rise in recall.

Table 10 shows the retrieval results for cross-concordances mapping terms within the same discipline, whereas table 11 shows the results for 2 interdisciplinary cross-concordances:

|       | Retrieved | Relevant | Rel_ret | Recall | Precision | P10    | P20    |
|-------|-----------|----------|---------|--------|-----------|--------|--------|
| FT    | 163.8     | 115.8    | 60.2    | 0.5934 | 0.5025    | 0.4635 | 0.4090 |
| FT-CK | 244.9     | 115.8    | 77.1    | 0.7096 | 0.4449    | 0.3826 | 0.3681 |
|       |           |          |         | 19.6   | -11.5     | -17.5% | -10.0% |

Table 10. Test 2 evaluation results for intradisciplinary cross-concordances (N=6)

|       | Retrieved | Relevant | Rel_ret | Recall | Precision | P10    | P20    |
|-------|-----------|----------|---------|--------|-----------|--------|--------|
| FT    | 129.9     | 78.2     | 44.3    | 0.6303 | 0.3129    | 0.2500 | 0.2459 |
| FT-CK | 332.4     | 78.2     | 59.8    | 0.7805 | 0.2388    | 0.1333 | 0.1292 |
|       |           |          |         | 23.8%  | -23.7%    | -46.7% | -47.5% |

Table 11. Test 2 evaluation results for interdisciplinary cross-concordances (N=2)

Contrary to the analysis for test 1, the differences between intradisciplinary and interdisciplinary cross-concordances are not as big. For both sets, recall still increases over a simple free-text search whereas precision drops. Recall slightly increases for interdisciplinary cross-concordances whereas precision drops much lower. This might be due to the fact that only 2 interdisciplinary cross-concordances where evaluated, which could skew the results and might not show enough of a trend.

The results of the free-text search experiment could probably be much improved if the controlled terms and natural language terms were better integrated in the query formulation instead of just appending one to the other. One possibility would be to (automatically) translate the natural language terms into controlled vocabulary terms a-priori so that there is a better chance for mapping to different controlled vocabularies and databases (see Mayr et al., 2008 for one approach).

In conclusion, the information retrieval experiments show the positive effects of cross-concordances for search in heterogeneous databases. The retrieval results improve for all cross-concordances, however, interdisciplinary cross-concordances cause a higher (positive) impact on the search results. For all cross-concordances in both test scenarios, more relevant documents were found compared to the query types without the use of cross-concordances; in particular cases, the retrieved set was even more precise (increase in precision as well).

## 6. Conclusions and Outlook

After showing that leveraging cross-concordances for search can have a positive impact on the search results, we plan on implementing them in the vascoda portal. Already, we have utilized many of the cross-concordances for search in the German Social Science Information Portal sowiport[13], which offers bibliographical and other information resources (incl. 15 databases with 10 different vocabularies and about 2.5 million bibliographical references).

The implementation in sowiport, which, like the experiments, only uses the equivalence relations, looks up search terms in the controlled vocabulary and then automatically adds all equivalent terms from all available mapped vocabularies to the query. It is therefore similar to the methodology applied in experiment 2 (free-text search). Boolean commands function as separators between query phrases, that is, they remain intact after query expansion (i.e. each query part gets expanded separately). The term mapping is automatic and invisible to the searcher, a small icon symbolizes the transformation (clicking on the icon lists the appended query terms).

---

[13] http://www.sowiport.de/

For further research and storage, a relational database was created to store the cross-concordances. To search and retrieve terminology data from the database, a web service (called heterogeneity service, see Mayr & Walter, 2008) was built to support cross-concordance searches for individual start terms, mapped terms, start and destination vocabularies as well as different types of relations. However, the database can also be queried on its own. The terminology mapping data as well as the web service can be made available for research purposes. Some mappings are already in use for the domain-specific track at the CLEF (Cross-Language Evaluation Forum) retrieval conference (Petras, Baerisch & Stempfhuber, 2007). Other features and applications of cross-concordances like switching, interacting or manipulating will be explored in later research.

Another option for storing and querying is the semantic web-based SKOS standard (Simple Knowledge Organization System)[14]. The goal of the SKOS standard, which is in W3C working draft status, is to formulate a standard that supports the use of controlled vocabularies for implementation in semantic web applications. The draft contains a section on mapping vocabularies. Once the SKOS standard is stabilized, we will make our mapping data available in this format.

One interesting area is the creation of mappings through a pivot vocabulary for resources where no means are available for direct mapping. If vocabulary A is mapped to vocabulary B and B is mapped to vocabulary C, it might be possible to create a mapping A → C by using the mapping information through the pivot vocabulary B. We hope that with the development of a standard for presentation and exchange, more mappings and vocabularies become available for further research.

## Acknowledgements

The project was funded by BMBF, grant no. 01C5953. We wish to thank all our project partners for their collaboration. We especially appreciate the help of our colleagues Anne-Kathrin Walter and Stefan Baerisch, who implemented most of the technical infrastructure and helped with the evaluation.

---

[14] http://www.w3.org/2004/02/skos/